\documentclass[useAMS,usenatbib]{mn2e}

\usepackage{graphicx}

\newcommand\hi{\mbox{H\,{\sc i}}}

\title[A Wide-Field Study of \mbox{Holmberg\,II}]
      {A Deep, Wide-Field Study of \mbox{Holmberg\,II} with Suprime-Cam:
      Evidence for Ram Pressure Stripping%
       \thanks{Based on data collected at Subaru Telescope, which is
       operated by the National Astronomical Observatory of Japan.}}
\author[Edouard J. Bernard et al.]
       {Edouard J. Bernard,$^{1}$\thanks{E-mail: ejb@roe.ac.uk}
        Annette M. N. Ferguson,$^{1}$ Michael K. Barker,$^{1}$ \newauthor
        Michael J. Irwin,$^{2}$ Pascale Jablonka,$^{3,4}$ Nobuo Arimoto$^{5,6}$\\
$^{1}$SUPA, Institute for Astronomy, University of Edinburgh,
    Royal Observatory, Blackford Hill, Edinburgh EH9 3HJ\\
$^{2}$Institute of Astronomy, Cambridge University, Cambridge CB3 0HA\\
$^{3}$Laboratoire d'Astrophysique, Ecole Polytechnique F\'ed\'erale de Lausanne
    (EPFL), Observatoire, CH-1290 Sauverny, Switzerland\\
$^{4}$GEPI, Observatoire de Paris, CNRS UMR 8111, Universit\'e Paris Diderot,
    F-92125 Meudon Cedex, France\\
$^{5}$Subaru Telescope, 650 North A'ohoku Place, Hilo, Hawaii 96720, U.S.A.\\
$^{6}$Graduate University for Advanced Studies, 2-21-1 Osawa, Mitaka, Tokyo
    181-8588, Japan}

\begin{document}

\date{Accepted ?. Received 2012 August 23; in original form 2012 July 25}

\pagerange{\pageref{firstpage}--\pageref{lastpage}} \pubyear{2012}

\maketitle

\label{firstpage}

\begin{abstract}
 We present a deep, wide-field optical study of the M81 group dwarf
 galaxy Holmberg~II (HoII) based on Subaru/Suprime-Cam imaging.
 Individual stars are resolved down to $I\sim25.2$, i.e. about
 1.5~mag below the tip of the red giant branch (RGB). We use resolved
 star counts in the outskirts of the galaxy to measure the radial
 surface brightness profile down to $\mu_V\sim 32$ mag arcsec$^{-2}$,
 from which we determine a projected exponential scalelength of
 $0.70\arcmin\pm0.01\arcmin$ (i.e. $0.69\pm 0.01$~kpc). The
 composite profile, ranging from the cored centre out to R=7$\arcmin$,
 is best fit by an EFF profile which gives a half-light radius of
 $1.41\arcmin\pm0.04\arcmin$ (i.e. $1.39\pm 0.04$~kpc), and an
 absolute magnitude M$_V=-$16.3. The low surface-brightness stellar
 component of HoII is regular and symmetric and has an extent much
 smaller than the vast \hi\ cloud in which it is embedded. We
 compare the spatial distribution of the young, intermediate age,
 and old stellar populations, and find that the old RGB stars are
 significantly more centrally concentrated than the young stellar
 populations, contrary to what is
 observed in most dwarf galaxies of the Local Universe. We discuss
 these properties in the context of the comet-like distribution of
 \hi\ gas around HoII, and argue for the presence of a hot intragroup
 medium in the vicinity of HoII to explain the contrasting
 morphologies of the gas and stars.
\end{abstract}


\section{Introduction}\label{sec:1}

In spite of their intrinsic faintness and minimal contribution to the
total light, the stellar outskirts of galaxies hold crucial
information about the processes of galaxy formation and evolution.
Interactions, mergers and accretions all leave their imprint on the
outer stellar populations in the form of substructures, streams and
diffuse haloes. In addition, the long evolutionary timescales and high
sensitivity to external influences means that coherent substructures
are easier to detect and longer-lived than in the denser inner
regions.

Considerable effort has been devoted to studying the outer regions of
massive galaxies \citep[see, e.g.][and references therein]{bar12}
which, according to hierarchical models of galaxy formation, have
acquired a significant fraction of their mass through mergers and
accretion episodes.  The stellar peripheries of low mass dwarf
galaxies have been much less studied. Most work to date has focused on
Local Group (LG) galaxies (e.g. Fornax: \citealt{col05}; NGC6822:
\citealt{deb06}; Sculptor: \citealt{wes06}),
although a few dwarfs in nearby groups have also been targeted
\citep[e.g.][]{rys11}.  The strong morphology-density relation
exhibited by dwarf galaxies suggests external mechanisms play a major
role in shaping their evolution \citep[e.g.][]{wei11}; surveying their
stellar outskirts may therefore yield clues on the dominant processes
involved.

The global picture emerging from previous work is that most dwarf
galaxies have a smooth, and generally old and metal-poor `halo'
surrounding a more concentrated, younger
and/or more metal-rich population.  While the stellar haloes of large
galaxies are believed to have formed from the accretion of smaller
galaxies at high redshift, it is unclear whether this process has been
significant in dwarfs.  For example, pure accretion cannot explain the
main properties of dwarf galaxies' haloes, i.e., smooth distribution,
and the existence of age and/or metallicity gradients. Instead, these
observations suggest either a `shrinking' scenario, in which the
region of active star formation contracts as gas supply diminishes
\citep{hid09,zha12}, or radial migration of stars formed close to the
centre towards the outskirts \citep{sti09}.

Here we analyse the stellar outskirts of the dwarf galaxy Holmberg\,II
(hereafter abbreviated as HoII).  HoII is a dwarf irregular galaxy in
the M81 group that was discovered by \citet{hol50} while surveying the
galaxies in this group. Due to its location on the near-side of the
group ((m$-$M)$_0$=27.65, i.e. 3.4~Mpc), as well as its proximity to
the Sc spiral galaxy NGC2403 and similar radial velocity, it is
usually associated with the NGC2403 subgroup along with three other
dwarf irregular galaxies \citep{kar02}.
HoII is very similar to the Small Magellanic Cloud (SMC) in terms of
absolute magnitude, \hi\ and total mass: M$_B\sim-$16.7, M$_{\rm HI}
\sim 6 \times 10^8 M_{\sun}$, and M$_{\rm tot} = 2.1 \times 10^9
M_{\sun}$ \citep{wal07,oh11} compared to M$_B\sim-$16.1, M$_{\rm HI}
\sim 4 \times 10^8 M_{\sun}$, and M$_{\rm tot} = 2.4 \times 10^9
M_{\sun}$ for the SMC \citep{sta99,sta04}.

 \defcitealias{bur02}{BC02}

\begin{figure}
 \includegraphics[width=8.3cm]{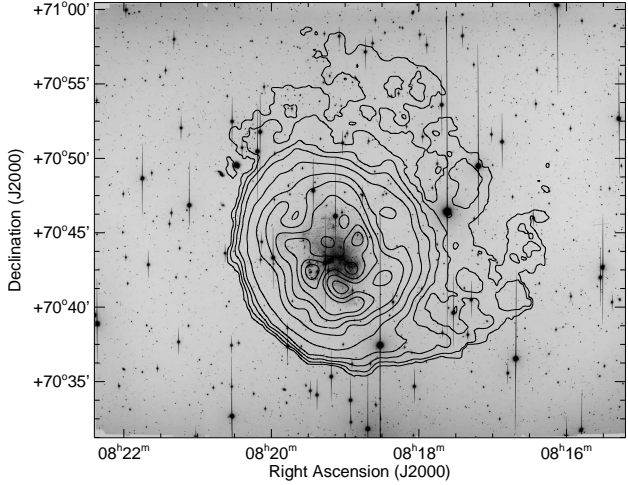}
 \caption{Mosaic V image of HoII, showing the whole field-of-view of
   our Subaru data ($\sim$35.8$\arcmin\times$29.3$\arcmin$). The \hi\
   contours from \citetalias{bur02}, ranging from $N_{HI}= 0.1$ to
   $19\times 10^{20}$ atoms cm$^{-2}$, are overplotted. Note the
   comet-like morphology of the outermost contour.\label{fig1}}
\end{figure}

While HoII has been observed at all wavelengths, to date the only deep
resolved stellar populations study comes from HST/ACS observations
\citep{wei09} which cover a relatively small fraction of the galaxy.
The need for deep wider data is especially motivated by the striking
morphology of its \hi\ cloud. From deep VLA data, \citet[hereafter
BC02]{bur02} found that the distribution of neutral hydrogen has a
cometary appearance -- compressed on one side with a faint extended
component on the opposite side -- with the tail pointing away from the
centre of the M81 group. \citetalias{bur02} argued that HoII is moving
toward the M81 group and that ram pressure from a hot intragroup
medium (IGM) is responsible for the \hi\ morphology, although they
could not rule out the alternative interpretation of a gravitational
interaction between HoII and one of its fainter neighbours.

\begin{figure*}
 \includegraphics[width=15.0cm]{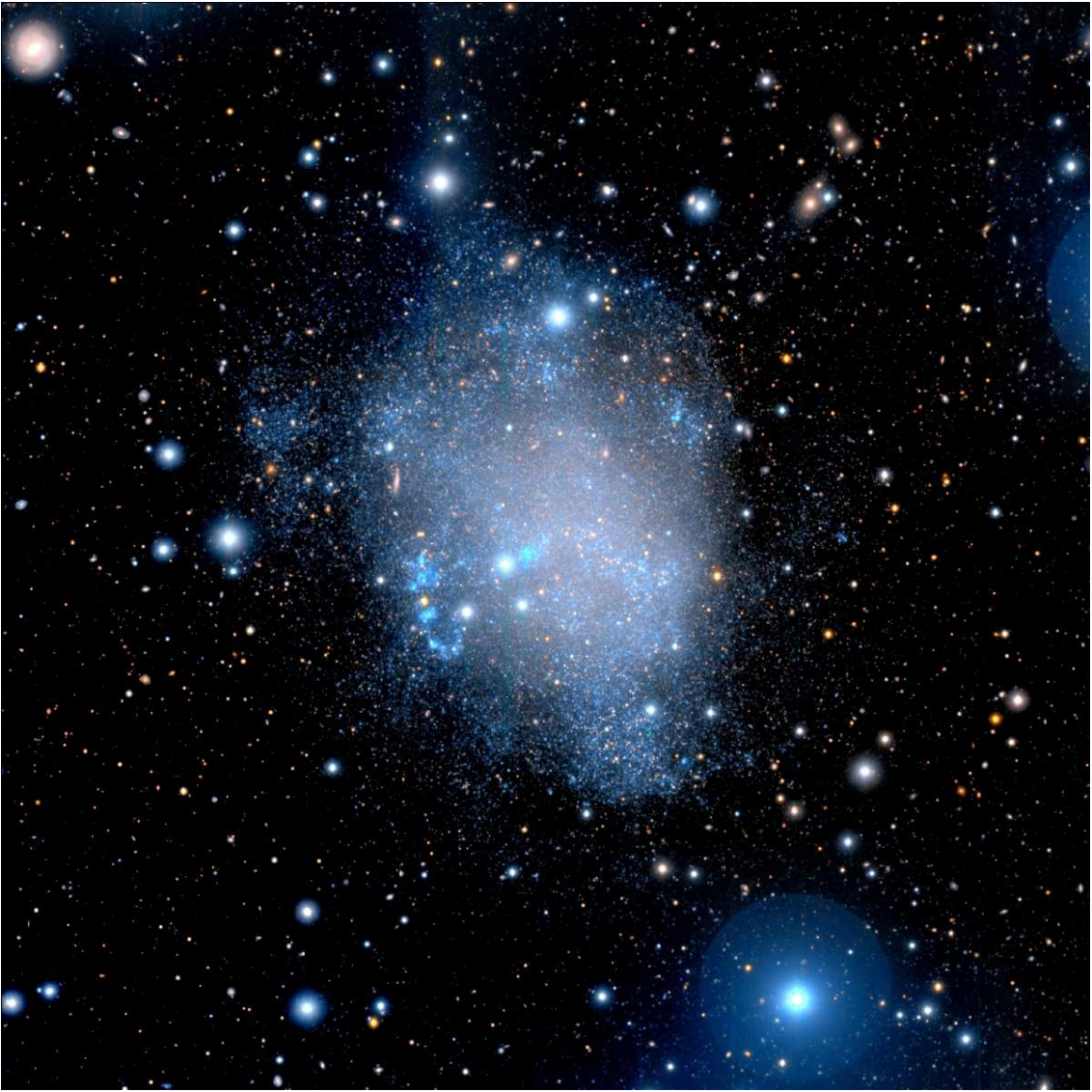}
 \caption{Color composite mosaic of HoII from our Subaru data. The image is
 cropped to $\sim$14$\arcmin$ on a side; North is up and East to the left.
 \label{fig2}}
\end{figure*}

Possible signatures of ram pressure stripping have been observed in a
number of galaxies of the local universe
\citep[e.g.][]{con83,gav95,ryd97,ken04,chu07,mcc07}. However, most of
these galaxies are too distant to be resolved into individual stars
and their analyses have been confined to fairly high surface
brightness inner regions. Knowledge of the distribution of resolved
stellar populations at very large radii is fundamental because the
stars do not respond to ram pressure. On the other hand, tidal forces
affect gas and stars equally so both components should exhibit similar
asymmetries.  Comparing the large-scale distribution of the stars with
that of the gas in HoII has the potential to reveal whether the \hi\
morphology was caused by ram pressure, tidal forces, or a combination
of both.

In this paper we present a deep, wide-field study of HoII based on
Subaru/SuprimeCam data, and analyse the properties of its stellar
populations in the context of the comet-like shape of the \hi\ cloud
in which it is embedded.  In Section~\ref{sec:2}, we describe the
observations and data reduction, and present the resulting CMDs in
Section~\ref{sec:3}. The spatial distribution of the various stellar
populations are decribed in Section~\ref{sec:4}. In
Section~\ref{sec:5}, we present the radial profile and constrain the
spatial extent of HoII. We discuss the implications of our results
regarding the peculiar stellar and \hi\ distributions in
Section~\ref{sec:6}, and summarize the main results in
Section~\ref{sec:7}.

\section{Observations and Data Reduction}\label{sec:2}

\subsection{Observations and Image Processing}\label{sec:2.1}

The observations of HoII (=UGC~4305 =DDO~50 =Arp~268) were obtained in
service mode with the Suprime-Cam instrument \citep{miy02} on the 8-m
Subaru telescope (P.I.: M. Barker). The data were acquired on the
night of December 18, 2009. Suprime-Cam is a mosaic camera made of 10
chips disposed in two rows of five chips -- thus producing 10 images
per exposure -- leading to a large field-of-view (FOV;
34$\arcmin\times$27$\arcmin$). A single pointing was therefore
sufficient to cover the relatively small galaxy
\citep[$R_{25}=4.1\arcmin$; from HYPERLEDA:][]{pat03}, including the
large \hi\ cloud in which it is embedded \citepalias[$R\sim
16\arcmin$;][]{bur02}. This is illustrated in Figure~\ref{fig1},
showing the V-band mosaic from our Subaru data where the \hi\ contours
from \citetalias{bur02} have been overlaid.

Ten exposures were obtained in each band, with individual exposure
times of 600s in Johnson $V$ and 240s in Cousin $I$. A small dithering
pattern was used to cover the gaps between the chips as well as limit
the effects of bad pixels and other camera defects, resulting in a
mosaic image of $\sim$35.8$\arcmin\times$29.3$\arcmin$. The
observations were carried out in photometric conditions, with seeing
in the range $0.65-0.95\arcsec$. However, due to HoII's location at
high declination, it can only be observed at relatively large
airmasses from Hawaii ($\ga$1.6), so the median seeing measured in the
$V$ and $I$ images is 0.96 and 0.76$\arcsec$, respectively.

The image processing procedures were similar to those followed by
\citet{bar09,bar12}. After converting each exposure to a single
multi-extension FITS file, all images and calibration frames were run
through a variant of the data reduction pipeline developed for
processing Wide Field Camera (WFC) data from the Isaac Newton
Telescope (INT) -- for futher details see
\citet{irw85,irw97,irw01,irw04}.

First stage image processing steps included bias and
overscan-correction, together with trimming to the reliable active
detector area. Master flats were created by stacking a dithered set of
10 $V$ and 12 $I$ twilight sky exposures. The flat-fielding step also
corrects for internal gain variations between the detectors. After
flat-fielding the dark sky $I$-band images were examined for signs of
fringing, but as for other similar Subaru data, this was found to be
negligible.

Prior to stacking, detector-level catalogues were generated for each
individual processed science image to refine the astrometric
calibration and also to assess the data quality. For astrometric
calibration, a Zenithal polynomial projection \citep{gre02} was used
to define the World Coordinate System (WCS). A fifth-order polynomial
includes all the significant telescope radial field distortions
leaving just a six-parameter linear model per detector to completely
define the required astrometric transformations. The Two Micron All
Sky Survey (2MASS) point-source catalogue \citep{cut03} was used for
the astrometric reference system.

During the
stacking process the individual Subaru catalogues were used in
addition to the standard WCS solution, to further refine to the
sub-pixel level the alignment of the component images. The common
background regions in the overlap area from each image in the stack
were used to compensate for sky variations during the exposure
sequence and the final stack included seeing weighting, confidence
(i.e. variance) map weighting and clipping of cosmic rays.

As a final image processing step, catalogues were derived from the
deep stacks for each detector and their WCS astrometry was updated
prior to forming the mosaic over all detectors. Any residual small
offsets in the underlying sky level between each detector stack were
removed iteratively by visual inspection of a 4$\times$4 blocked
mosaic of the whole field.

Full resolution mosaics were used to provide an initial set of
full-field catalogues based on standard aperture photometry
\citep[e.g.][]{irw04}.
Since no standard star field was observed, we used INT WFC $V$- and
$i$-band observations centred on HoII and taken in photometric
conditions during April 2009 to derive the photometric calibration
for the mosaics. The INT WFC $V$,$i$ photometry was first
converted to the Johnson-Cousins system \citep[e.g.][]{mcc03} and
then directly used to calibrate the Subaru data with an estimated
systematic error of $\pm$2 per cent.

Figure~\ref{fig2} shows a colour composite image obtained
from the $V$ and $I$ mosaics, which highlights the irregular morphology
and strong contribution from the young, blue supergiant stars.

\begin{figure}
 \includegraphics[width=8.0cm]{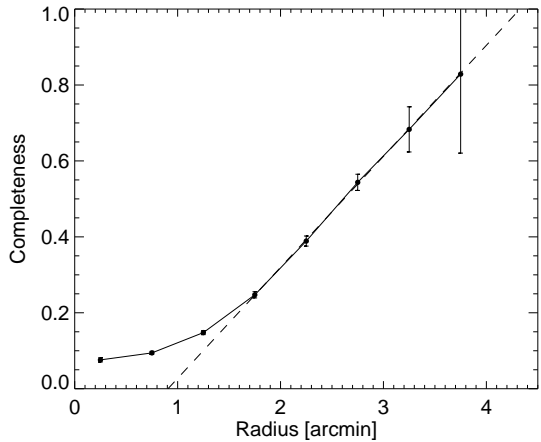}
 \caption{Completeness of the Subaru data, relative to the HST/ACS data,
 as a function of galactocentric radius for stars brighter than $I_0$=24.5.
 The  dashed lines is a fit to the points between R=1.5 and 4$\arcmin$.
\label{fig3}}
\end{figure}

\subsection{Profile-fitting Photometry}\label{sec:2.2}

Given the very high stellar density in HoII, we subsequently decided
to perform profile-fitting photometry which is better suited to obtain
accurate measurements in crowded fields. The PSF photometry was
carried out with the {\sc daophot/allstar/allframe} suite of programs
\citep{ste94} as follows. We performed a first source detection at the
3-$\sigma$ level on the individual images, which was used as input for
aperture photometry. From these catalogues, 300 bright, non-saturated
stars per image were initially selected as potential PSF stars. An
automatic rejection based on the shape parameters was used to clean
the lists, followed by a visual inspection of all the stars to remove
the remaining unreliable stars. We ended up with clean lists
containing at least 150 good PSF stars per image. Modelling of the
empirical PSF with a radius of 12 pixels was done iteratively with
{\sc daophot}: the clean lists were used to remove all the stars from the
images except PSF stars, so that accurate PSFs could be created from
non-crowded stars. At each iteration, the PSF was modelled more
accurately and thus the neighbouring stars removed better. Every 3 to
5 iterations, the degree of PSF variability across the image was also
increased, from constant to linear, then quadratically variable.

The following step consisted of profile-fitting photometry on the
individual images using {\sc allstar} with the empirical PSFs previously
created. The resulting catalogues were matched on a chip-by-chip
basis, keeping only the objects for which the PSF fitting converged in
at least three images per band to limit the number of false
detections, to create one clean stellar catalogue per chip.  These
catalogues were further cleaned by rejecting extended objects based on
the sharpness parameter, then merged to produce a master catalogue for
the whole Suprime-Cam FOV. This catalogue was then used as
the input star list for {\sc allframe}, which was run on all the individual
images at once. The output of {\sc allframe} consists of a catalogue of PSF
photometry for each individual image. A robust mean magnitude was
obtained for each star by matching these catalogues.

The final photometry was calibrated to the Johnson--Cousin standard
system by matching $\sim$1750 bright stars in common with the
calibrated aperture photometry described in the previous Section. The
uncertainty on the offset between the aperture and PSF photometry is
smaller than 0.001 in both bands and therefore represents a negligible
contribution to the total magnitude uncertainties.

According to the \citet{sch98} reddening maps, the area covered by our
observations suffers from minor foreground differential reddening,
with E(B$-$V) ranging from about 0.023 to 0.033. To obtain accurate
photometry over the whole FOV, each individual star was corrected
for reddening based on its location. We note that the corresponding dust
mask map indicates that a small area centred on HoII (out to
r$\sim$6$\arcmin$; i.e. $\sim$10 per cent of the Subaru FOV), as an
extragalactic source, was removed from the dust map and replaced with
the median value of the surrounding pixels. This means that the
star-by-star correction does not include extinction internal to HoII.

Finally, the astrometric calibration for the whole stellar catalogue
was obtained with the IRAF tasks {\it ccxymatch}, {\it ccmap}, and
{\it cctran} using $\sim$800 stars in common with Version 2.3.2 of the
Guide Star Catalog~II \citep{las08}. The accuracy of the resulting
astrometry is about 0.3$\arcsec$.

\subsection{Completeness}\label{sec:2.3}

To estimate the completeness of our data, we retrieved the HST/ACS
photometry of HoII from the ACS Nearby Galaxy Survey Treasury program
\citep[ANGST;][]{dal09}. While it only covers a tiny fraction of our
FOV ($\sim$2 per cent), it is located on the highest density area
which is the most affected by incompleteness. It is $\sim$3~mag deeper
than our Subaru photometry and much less affected by stellar crowding
thanks to the higher spatial resolution of the instrument. Therefore,
in the following we assume that it is 100 per cent complete in the
range of magnitudes covered by our Subaru data, and representative of
the intrinsic photometric properties of the stars in HoII.

We used {\sc daomatch} and {\sc daomaster} \citep{ste93} to match the
HST and Subaru photometric catalogues. These programs are based on the
robust `matching triangles' technique, rather than assuming a given
matching radius. This ensures a proper matching of the stars even in
very crowded regions, regardless of the possible translations,
rotations, scale changes, or flips of the coordinate systems.

Using all the stars from the area in common between the two
photometric catalogues, we find the 50 per cent completeness limits of
the Subaru data at $V_0$=23.93 and $I_0$=23.00.  However, the centre
of the galaxy is more affected by crowding than the outer regions, so
completeness varies significantly with radius.  To correct the radial
density profile shown in Section~\ref{sec:5}, which uses all the
stars with $I_0<$~24.5, we therefore estimate the completeness down to
this magnitude as a function of radius. The result is shown in
Figure~\ref{fig3}. For each annulus, the errorbar was calculated
by adding in quadrature the Poisson uncertainties on the number of
stars in the HST and Subaru subsamples. We find that completeness
varies from 10 per cent in the inner 0.5$\arcmin$ to 90 per cent at
R$\sim$4$\arcmin$. The dashed line is a fit to the points between
R=1.5 and 4$\arcmin$, and suggests that we can assume $\sim$100 per
cent completeness beyond 4$\arcmin$.

\section{Colour-Magnitude Diagrams}\label{sec:3}

\begin{figure}
 \includegraphics[width=8.3cm]{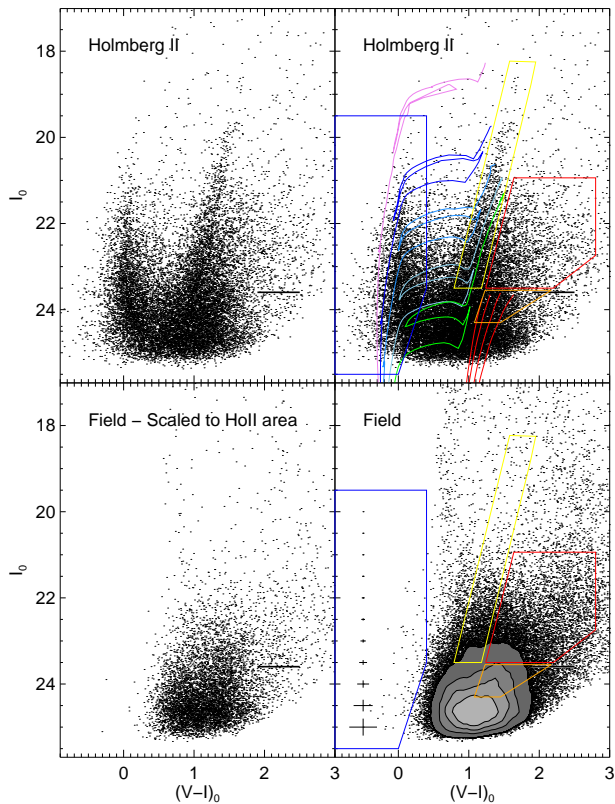}
 \caption{Extinction-corrected color-magnitude diagrams of HoII (top panels) and
 field stars (bottom panels). In the bottom right panel, the contour levels
 correspond to [6,12,18,24,30]$\times 10^3 {\rm \ stars \ mag}^{-2}$, and the
 error bars show the mean photometric errors as a function of magnitude for both
 field and HoII stars. The boxes used to select the various stellar populations
 are shown in the right panels: MS+BSG (blue), RSG (yellow), AGB (red), and RGB
 (orange). Selected isochrones from the Padua library \citep{mar08} have been
 overplotted in the top right panel (see text for details).
 \label{fig4}}
\end{figure}

The resulting colour-magnitude diagrams (CMDs) for the field and HoII
stars are shown in Figure~\ref{fig4}.  We used the 0.1 and 2$\times
10^{20}\rm{\ atoms\ cm}^{-2}$ \hi\ contours to separate the two
populations (see Section~\ref{sec:4}): stars outside the former are
assumed to be field stars while stars inside the latter are considered
HoII. The CMDs were cleaned of non-stellar objects using the
photometric parameters given by {\sc allframe}, namely $\sigma_{V,I}
\leq$~0.2 and $|${\tt SHARP}$|\leq$~1.  To enhance the features at
faint magnitudes, the CMD of the top left panel was further cleaned
using tighter constraints on the sharpness ($|${\tt SHARP}$|\leq$~0.5)
but is shown only for illustrative purpose. It contains $\sim$16,600
stars, instead of the $\sim$27,100 stars in the CMD of the top right
panel.

At $I_0$ brighter than $\sim$22 the field CMD, shown in the bottom
right panel of Figure~\ref{fig4}, harbors two prominent vertical
sequences: the main-sequence turn-off stars of the Milky Way (MW) halo
at ($V-I$)$_0\sim$~0.7, and dwarf stars in the MW disc at
($V-I$)$_0\sim$~2.5. At fainter magnitudes, these sequences are
overwhelmed by the contribution of unresolved background galaxies. We
note, however, that the HoII CMDs in the top panels correspond to an
area about 10 times smaller than that covered by the field stars, so
the effect of the contamination by foreground stars and background
galaxies is not as significant. To illustrate this, the bottom left
panel shows the CMD of a field with the same area as HoII.
The scaled field CMD contains $\sim$7600 objects.

To help identify the features in the HoII CMDs, in the top right panel
of Figure~\ref{fig4} we overplot isochrones from the Padua stellar
evolution library \citep{mar08}, shifted to the distance of HoII.  The
young isochrones have Z=0.002 and ages of 10, 20, 50, 100, and 160~Myr
from top to bottom, while the old ones shown in red are 12.5~Gyr old
with Z=0.0001, 0.0003, 0.001, and 0.002 from left to right.

 \begin{figure*}
 \includegraphics[width=12cm]{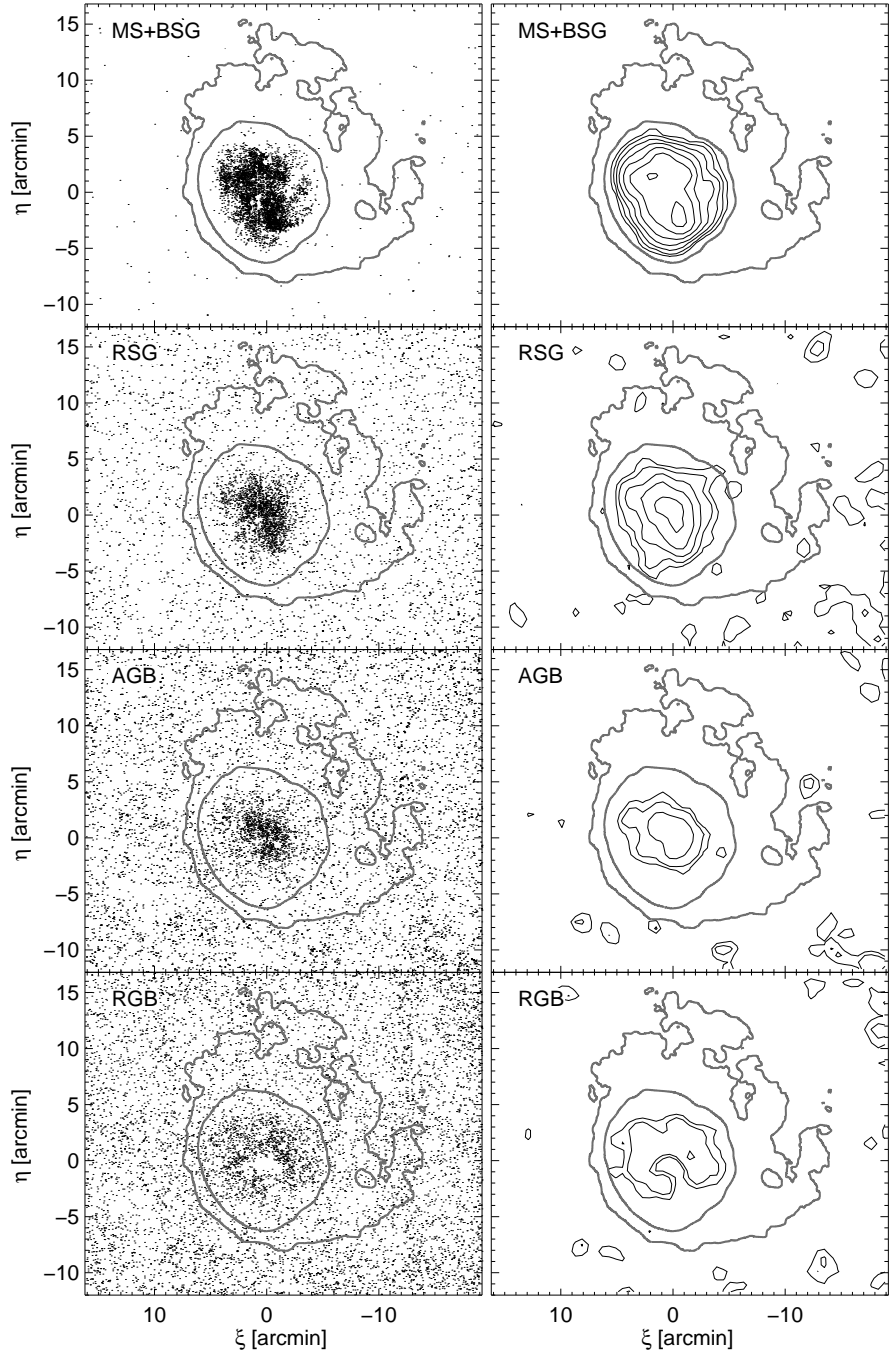}
 \caption{Spatial distribution of candidate MS+BSG, RSG, AGB and RGB stars in
 the whole Subaru FOV (left), and the corresponding contours (right).
 Stellar contours levels are arbitrary.
 The two \hi\ contours \citepalias[0.1 and 2$\times 10^{20}\rm{\ atoms\ cm}^{-2}$,
 from][]{bur02} used to separate HoII and field stars are shown as thick gray
 lines to use as guides.
 \label{fig5}}
\end{figure*}

The most prominent features of the HoII CMD are the blue and red
supergiant branches (BSG and RSG, respectively) of core helium-burning
stars at ($V-I$)$_0\sim0$ and 1.2, respectively. Their presence
indicates vigorous star formation in the past $\sim$160~Myr.  The BSG
branch curves toward the red below $I_0\sim24$, and joins the RSG
close to the red clump just below our limiting magnitude. Redward of
the RSG and at $I_0\ga 23.5$, a higher density of stars corresponding
to the red giant branch (RGB) is visible. The expected magnitude of
the tip of the RGB (TRGB), from \citet{kar02}, is shown as a thin line
at $I_0=23.60$. Some of the stars brighter than the TRGB and redder
than the RSG are genuine asymptotic giant branch (AGB) stars (see
Section~\ref{sec:4}), although they are difficult to separate from the
contamination of foreground stars and unresolved galaxies.  Finally, a
few dozen main sequence (MS) stars forming a diffuse band blueward of
the BSG branch are visible at about ($V-I$)$_0=-0.2$ and $I_0\ga 21.5$,
indicating star formation within the last 10~Myr. We note that while
some of these stars lie blueward of the youngest isochrones,
their concentration close to the centre of HoII confirms that they are
bona fide members of HoII. Their color spread, larger than expected
from photometric errors only, is a consequence of the crowding in this
region.

In the following Section, we analyse the spatial distributions of the
different stellar populations individually. They are selected on the
CMDs using the boxes shown in the right panels of
Figure~\ref{fig4}. The blue, yellow, red, and orange boxes enclose
the stars belonging to the MS+BSG ($\la$160~Myr), RSG (10--160~Myr),
AGB (a few hundred Myr to a few Gyr), and RGB ($\ga$1.5~Gyr),
respectively. For the MS+BSG selection, we chose to use the whole
range of luminosity since contamination by foreground and background
sources is virtually non-existent in this color range.

\begin{figure*}
 \includegraphics[width=10cm]{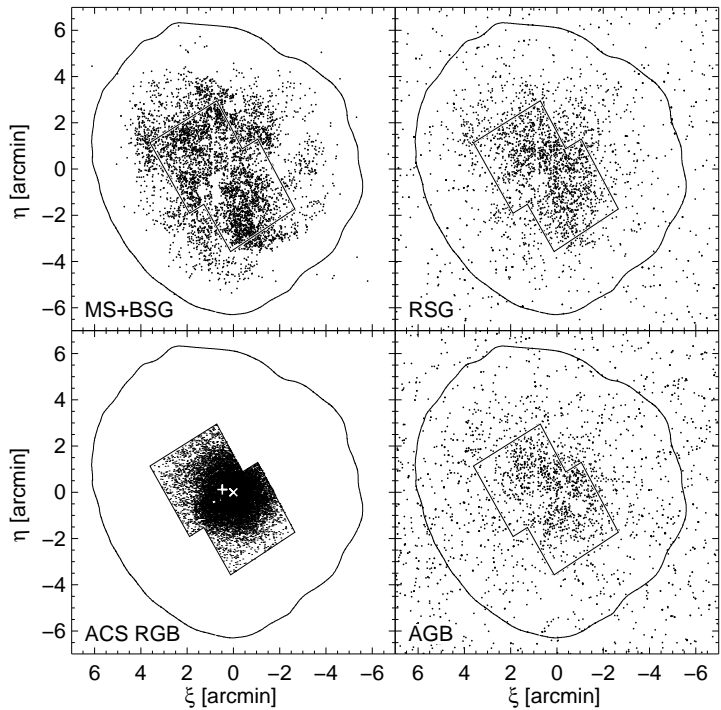}
 \caption{Zoomed-in spatial distribution of BSG, RSG, and AGB stars in the
 Subaru FOV, and RGB from HST/ACS data. Several small holes in the
 stellar distributions are due to highly saturated stars and their associated
 bleed spikes. The inner \hi\ contour of
 Figure~\ref{fig5}, as well as the outline of the combined footprint of
 the two ACS pointings, are shown. The cross and plus symbol in the bottom left
 panels represent the dynamical centre and the centre of the \hi\ contour,
 respectively, to highlight the offset between the two.
 \label{fig6}}
\end{figure*}

\section{Spatial Distribution}\label{sec:4}

In Figure~\ref{fig5} we show the spatial distribution of each
individual stellar populations -- selected using the boxes shown in
the left panels of Figure~\ref{fig4} -- and the corresponding
contours. From top to bottom, the panels show the distribution of the
MS+BSG, RSG, AGB, and RGB. To serve as guides, in each panel we also
plotted two \hi\ contours from the \citetalias{bur02} study,
representing densities of 0.1 and 2$\times 10^{20}\rm{\ atoms\
  cm}^{-2}$: the first one is their lowest density contour, while the
second is the innermost contour that includes most of the observed
MS+BSG stars belonging to HoII. We find that the shape of the latter
contour is close to an ellipse with radius R=6.0$\arcmin$, b/a=0.90,
and position angle of 27$\degr$. These are the contours we used to
separate HoII and field stars (see Section~\ref{sec:3}).

The top panels show that blue stars younger than about 160~Myr old are
confined to regions where the \hi\ density is higher than about
2$\times 10^{20}\rm{\ atoms\ cm}^{-2}$.  The very few objects outside
the contours are distributed uniformly over the FOV,
suggesting that they are either unresolved galaxies with similar
colors, or foreground blue HB, blue staggler, or white dwarf stars.
The outermost MS+BSG contour follows very closely the inner \hi\
contour, as expected if massive star formation only occurs above a
certain gas density threshold.  The other panels show that
contamination by field stars and unresolved galaxies increases
significantly at redder color and fainter magnitudes.  Nevertheless,
these maps do not show significant stellar concentrations outside of
the central contour, which suggests that most or all of the sources
outside this area are either foreground stars or unresolved galaxies.

\begin{figure*}
 \includegraphics[width=8.4cm]{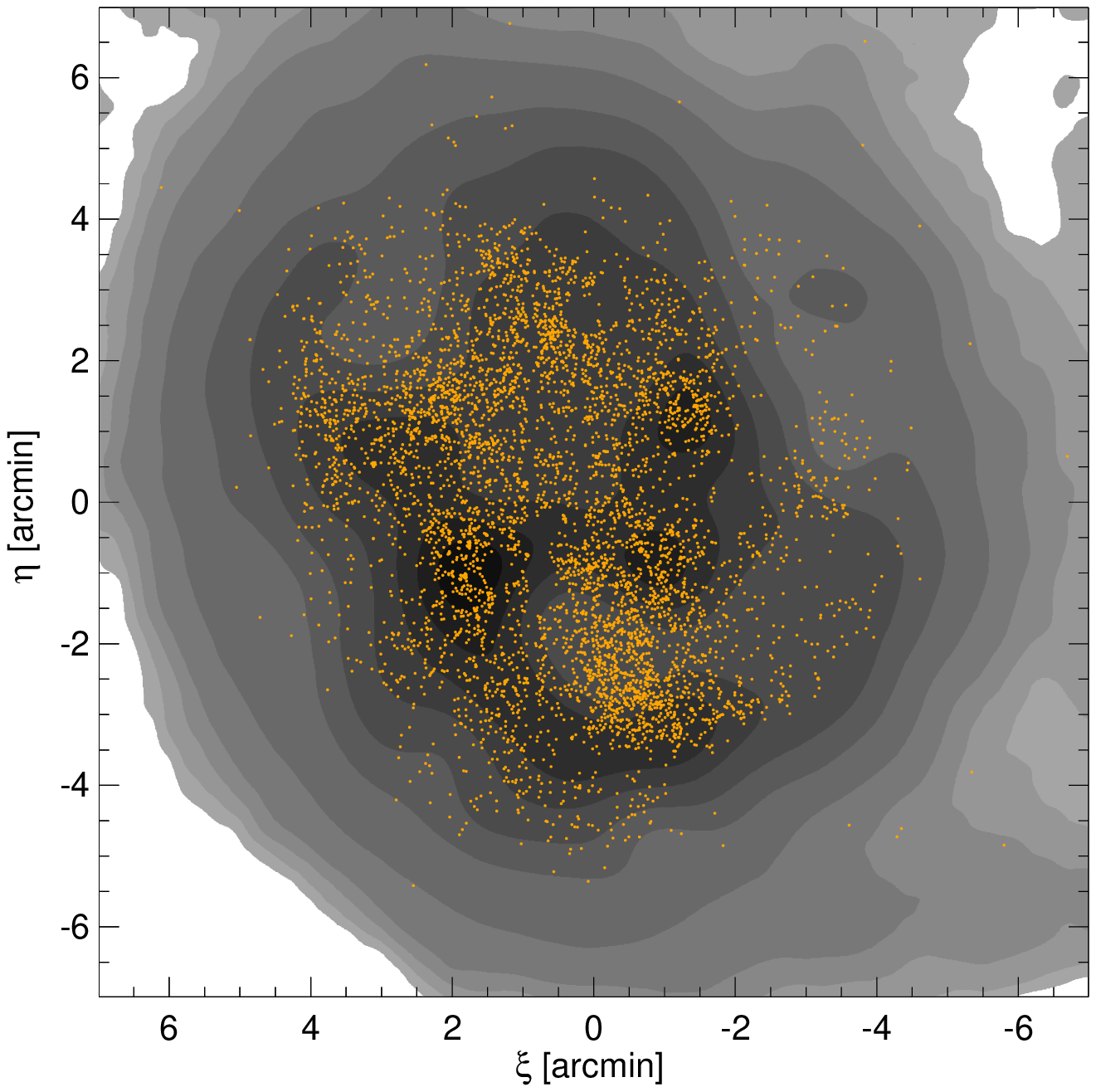}
 \includegraphics[width=8.5cm]{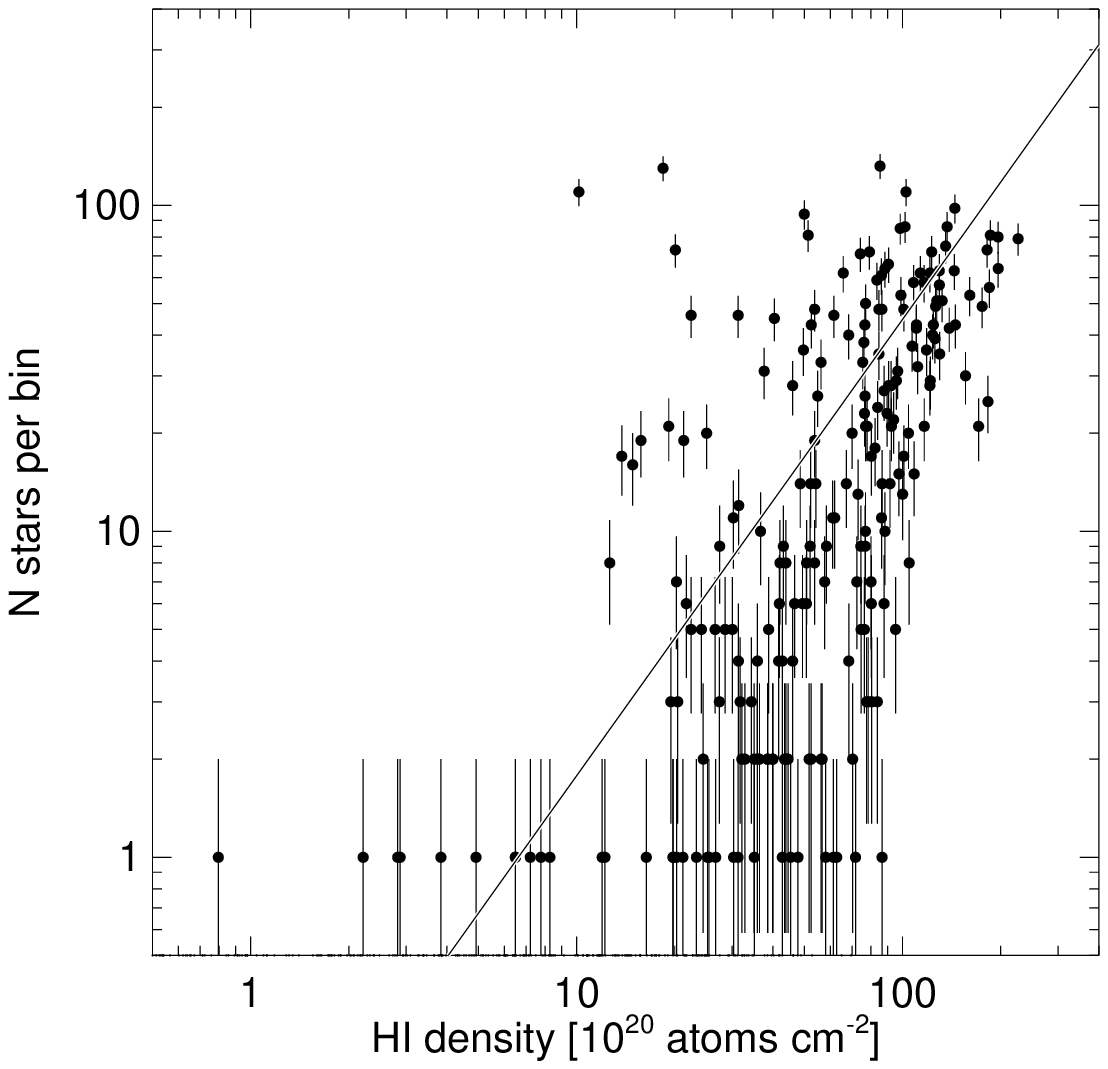}
 \caption{Left: spatial distribution of MS+BSG stars overlaid on the \hi\
 contours of \citetalias{bur02}. Right: Surface density of MS+BSG stars as
 a function of \hi\ density in square bins of 37.5$\arcsec$ on a side.
 The solid line represent the typical Kennicutt-Schmidt relation with a
 slope of 1.4 \citep{ken98}.
 \label{fig7}}
\end{figure*}

Interestingly, in Figure~\ref{fig5} the concentration seems to
increase from BSG to RSG stars, and from RSG to AGB stars, the
populations being more compact with increasing age. The RGB stars,
close to our detection limit, are the most affected by crowding which
explains the hole in the distribution at the centre of HoII. In
addition, some of the stars flagged as RGB may actually be RSG or AGB
stars that were shifted to the RGB box due to photometric errors.
Therefore, since the RGB distribution is not clear from our Subaru
data, we checked the distribution using the HST/ACS photometry of HoII
described in Section~\ref{sec:2.3}.
In the bottom left panel of Figure~\ref{fig6} we show the
distribution of RGB stars in the top 1.5~mag of the RGB from the HST
dataset. The MS+BSG, RSG, and AGB maps from the Subaru data are shown
in the other panels. In each panel, the black line outlines the
combined footprint of the two ACS pointings to facilitate comparison.

The difference between the distributions of the MS+BSG and RGB samples
is striking: despite the small area covered by the HST data, it is
obvious that the old, RGB stars are significantly more concentrated
than the younger stars, as already noted by \citet{bas11}. The former
are also distributed in a roughly circular distribution, whereas the
latter present a very irregular distribution. Within the ACS
footprint, the density of RGB stars almost vanishes close to the edges
while the density of blue stars is roughly uniform.  We note that the
distributions of Subaru and HST/ACS blue stars are very similar in the
area in common.
Using only the ACS data to avoid the uncertainties due to
incompleteness and different spatial coverage, we find exponential
scalelengths for the MS+BSG, RSG, AGB and RGB samples of
3.1$\arcmin\pm0.5\arcmin$, 1.7$\arcmin\pm0.3\arcmin$,
0.80$\arcmin\pm0.07\arcmin$, and 0.76$\arcmin\pm0.04\arcmin$,
respectively. Given the small FOV,
we did not apply a background correction to these profiles.

Figure~\ref{fig6} also shows that the blue stars, besides
having a larger spatial extent, also seem to delineate short spiral
arms.  This morphology is not due to highly saturated stars or
artifacts due to image defects, and completeness down to $I_0=$24.5 is
$\sim$100 per cent at large radii. These arms follow the same
counter-clockwise orientation as the \hi\ arms observed by
\citet{bur04} in the northwest of HoII, which further suggests they
are real.  The lack of similar arms in the distribution of AGB stars
suggests these are likely transient features.

The good agreement between the distribution of young stars and the
\hi\ gas is further illustrated in Figure~\ref{fig7}. The left
panel presents the distribution of MS+BSG stars overlaid on the \hi\
countours of \citetalias{bur02}, and shows that these are usually
located in the higher column density areas. In the right panel, we
plot the number of MS+BSG stars in bins of 37.5$\arcsec$ on a side as
a function of the \hi\ density in the same bins. Note that for this
panel we have used the higher resolution map of \citet[natural
weighting, beam size $\sim13\arcsec$]{wal08} to better sample the
scales of our stellar distribution. It shows that the distributions
of young stars and \hi\ gas are correlated. A few points at high gas
density and low star counts are artifacts due to the holes left by
very saturated stars and their bleed spikes. On the other hand, the
datapoints with high stellar and low gas density are real: most of
them are located in the main \hi\ hole at $\xi\sim0\arcmin$ and
$\eta\sim-2\arcmin$, where \citet{wei09} found elevated star formation
about 50~Myr ago, but only low levels in the past 10--20~Myr.
Interestingly, even though we use the number of MS+BSG stars (i.e.
younger than about 150~Myr) as a rough proxy for the star formation
rate, we find that the line representing the typical Kennicutt-Schmidt
relation with a slope of 1.4 \citep{ken98} provides a good match to
our data.

Another interesting characteristic in Figure~\ref{fig6} is the
apparent offset between the centres of the MS+BSG and HST/RGB star
distributions, shown in the bottom left panel as the open circle and
cross, respectively.  We note that we applied a small shift
($\la$4$\arcsec$) to the astrometry of the HST data to correct for a
slight mismatch between the two catalogues, so the offset is not due
to inaccurate astrometry.  We find that the dynamical centre listed in
\citet[$\alpha_{2000}=8^h19^m05.6^s$, $\delta_{2000}=70\degr 43\arcmin
25\arcsec$]{ste00} provides an excellent fit to the distribution of
RGB stars.
On the other hand, the centre of the inner \hi\ contour, which follows
very closely the distribution of MS+BSG stars, is offset from the
dynamical centre by 30$\arcsec$ (0.49~kpc) east and 7$\arcsec$
(0.1~kpc) north. While the origin of this shift is unclear, it is
reasonable to suspect that it is related to the processes that shaped
the outer \hi\ envelope.

  \defcitealias{deb08}{de Blok et al.\ 2008}
  \defcitealias{mcc07}{McConnachie et al.\ 2007}
  \defcitealias{oh11}{Oh et al.\ 2011}
  \defcitealias{wal08}{Walter et al.\ 2008}
  \defcitealias{gen12}{Gentile et al.\ 2012}

\section{Radial Profile}\label{sec:5}

In order to obtain the radial density profile of HoII, one needs a
reliable estimate of its ellipticity, position angle, and location of
its centre.  We use the dynamical centre described in the previous
Section, which is also the centre of the intermediate-age and old
stellar populations ($\ga$1.5~Gyr old). The large uncertainty on the
inclination of HoII \citepalias{bur02,deb08,wal08,oh11,gen12} prevents
us from properly constraining the morphology of the galaxy.  However,
given the very low ellipticity of the inner \hi\ contours
(1$-$b/a$\sim$0.1) and circular distribution of the RGB stars, we can
assume circular symmetry for the purpose of calculating the radial
profiles.

While resolved star counts are the optimal way to probe the very low
surface brightness structure of galaxies \citep[e.g.][]{bar09}, they
are not ideal at small radii due to crowding and incompleteness.  As
described in Section~\ref{sec:2.3}, crowding is particularly severe in
the central part of HoII.  However, it is possible to use diffuse
light in this area and combine the contribution of both to obtain a
composite profile extending over the whole radius.

For the diffuse light surface-brightness profile, we used the median
pixel value in concentric circles, after masking the saturated stars
and their associated bleed spikes, as well as bright background
galaxies.  The sky value was estimated from the mode of the pixel
value distribution of the entire image. The pixel values were then
converted to magnitudes per arcsec$^2$, calibrated using the same
zero-point as for the stellar photometry, and corrected for reddening
using E(B$-$V)=0.032 \citep{sch98}.

The same concentric circles were used for the star count profile, in
which we counted the number of stars brighter than $I_0$ = 24.5. This
limit was chosen as the best compromise between sufficient number
statistics and reasonable completeness correction. The background
contaminant level was estimated from the density of point sources in a
wide circular annulus in a region where HoII's radial profile is flat
within the uncertainties (7.5\arcmin$\leq\rm{r}\leq10.5\arcmin$). For
the area covered by the HST data, we applied the correction for
completeness determined in Section~\ref{sec:2.3}. As for the diffuse
light, the radial density profile $\Sigma(r)$ was then converted to a
magnitude scale using the relation $\mu(r)=-2.5\ {\rm log}\ \Sigma(r)
+ ZP$.  However, here the zero-point $ZP$ was estimated by matching
the overlapping region of the stellar and diffuse light profiles
(2.5\arcmin$\la\rm{r}\la4\arcmin$).

\begin{figure}
 \includegraphics[width=8.5cm]{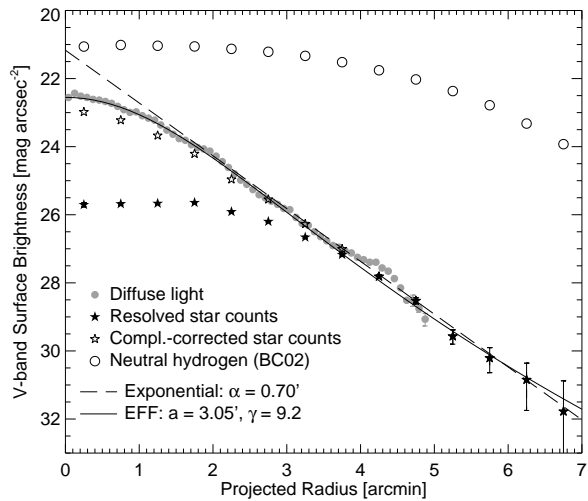}
 \caption{Background-subtracted surface brightness profiles from the diffuse
 light (gray filled circles), resolved star counts (filled stars), and
 completeness corrected star counts (open stars). The open circles represent the
 \hi\ density profile, from \citetalias{bur02}. The dashed and solid lines are
 exponential and EFF profiles fit to the composite surface brightness profile
 (see text).
 \label{fig8}}
\end{figure}

The resulting profiles are presented in Figure~\ref{fig8}: the
diffuse light and stellar surface-brightness profiles are shown as the
gray filled circles and black filled stars, respectively. The open
stars show the completeness-corrected star count profile.  Although
the completeness-corrected star counts and the diffuse light give
reassuringly similar information in the inner regions, the star count
data allow the profile to be extended to ${\rm R} \sim$~7$\arcmin$
where $\mu_V \sim$~32 mag arcsec$^{-2}$.

The surface brightness is well-described by an inner core (to
$\sim$1.5$\arcmin$) and a smooth outer decline.  We find that an EFF
\citep{els87} profile provides the best description over the entire
radial range, while an exponential fit also works beyond the inner
core.  On the other hand, the \citet{kin62} and \citet{plu11} profiles
(not shown) tend to under- and overestimate the stellar density beyond
$\sim$3$\arcmin$, respectively. The exponential fit to the profile
between R=1.5 and 7$\arcmin$ is shown as the dashed line, and has a
scalelength of $0.70\arcmin\pm0.01\arcmin$, or $0.69\pm 0.01$~kpc at
the distance of HoII.  The EFF fit (solid line) leads to a half-light
radius of $1.41\arcmin\pm0.04\arcmin$ (i.e. $1.39\pm 0.04$~kpc) and
absolute magnitude M$_V=-$16.3.

Our scalelength is slightly smaller than the estimates from
\citet[$h_B=0.99\arcmin$ and $h_R=0.86\arcmin$ after correcting for
the difference in ellipticity;][]{swa99}.  The discrepancy between the
two estimates is most likely due to a combination of the shallower
depth of their data ($\mu_R\la27.5$) and the steepening of our profile
at larger radii.  On the other hand, the radial profile of HoII shown
in \citet{oh11} extends further out than the one presented here (out
to R$\sim$9~kpc). We believe their extent may be overestimated as a
consequence of using a relatively large ellipticity (b/a~$\sim$~0.66),
and a position angle determined from the \hi\ data (PA=175$\degr$)
which is offset by about 35$\degr$ from the position angle of the
optical data \citep[PA=30$\degr$;][]{swa99}.

We thus find that the stellar component of HoII is significantly more
compact than the \hi\ cloud in which it is embedded, which extends out
to $\sim 16\arcmin$ \citepalias{bur02}. Even if some stars belonging
to HoII are present beyond ${\rm R} \sim$~7$\arcmin$, their
contribution to the galaxy luminosity is likely to be negligible.
Assuming an exponential profile, such a population would represent
less than 0.1 per cent of the total galaxy light.

\begin{figure*}
 \includegraphics[width=13.5cm]{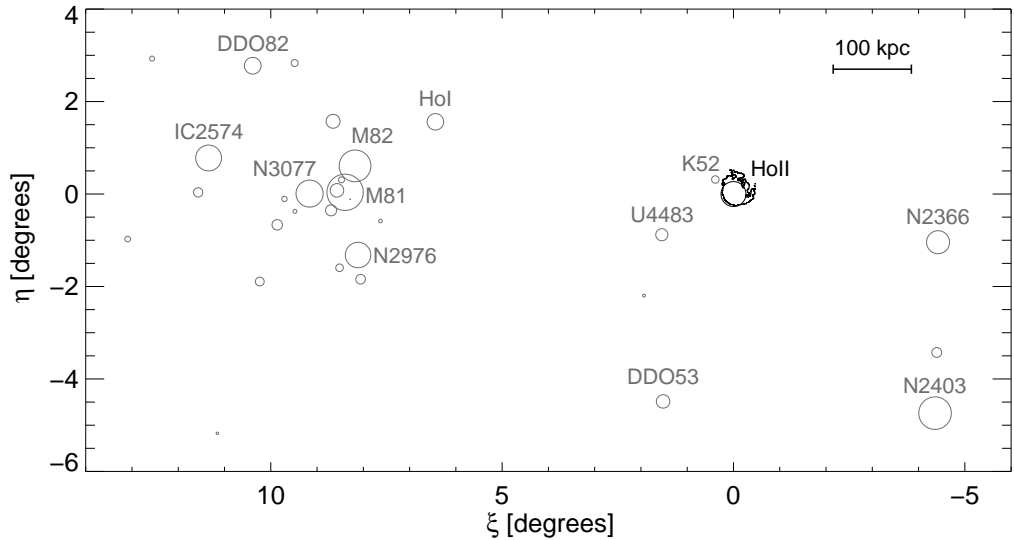}
 \caption{Distribution of galaxies in the M81 group in standard coordinates,
 where the symbol size is proportional to the apparent integrated B-band
 magnitude \citep{kar02}. HoII is shown as the black circle at (0,0), and the
 main galaxies of the group are labeled. The outermost \hi\ contour of HoII from
 \citetalias{bur02} is also shown, scaled up in size by a factor of 2.
\label{fig9}}
\end{figure*}

\section{Discussion}\label{sec:6}

\subsection{Global Structure}\label{sec:6.1}

Using the resolved stars counts in the outer regions of HoII revealed an
extended structure that is an order of magnitude fainter than what can
be reached with diffuse light. Outside the central 2$\arcmin$ area,
the profile is well-fit by an exponential model out to R=7$\arcmin$
with a scalelength of $0.69\pm 0.01$~kpc. However, we have also found
that the scalelength is different for each stellar population, in the
sense that it decreases with increasing age. This is contrary to what
is usually observed in dwarf galaxies in the nearby Universe
\citep[e.g.][]{zha12}, where the region of active star formation has
been shrinking with time.  This implies that HoII would have been a
more compact galaxy in the past.  The very low density of RGB stars in
the outskirts of the HST fields compared to that of the younger stars
can be interpreted as a significantly lower SFR in these parts
$\ga$1.5~Gyr ago (the minimum age of stars in the top magnitude of the
RGB) than in the past $\sim$160~Myr. This is in good agreement with
the SFHs obtained from these HST data, which show a higher SFR
starting about 300~Myr ago \citep[see also \citealt{mcq10}]{dal12}, as
well as a ``dramatic rise in the SFR over the past 50~Myr"
\citep{wei08}.

Given the large amount and relatively high density of the neutral
hydrogen surrounding HoII \citepalias{bur02}, it is somewhat
surprising that star formation was not more vigorous prior to about
300~Myr.  This could suggest that the extended \hi\ has recently been
acquired or that it previously had too low a density for star
formation to be significant. The process that shaped the spectacular
\hi\ envelope could have also increased the gas pressure in the inner
regions of the galaxy and triggered the large scale star formation
that we observe today. Two potential mechanisms to explain this are
ram pressure by a hot IGM and a tidal interaction with a companion
galaxy, which we discuss and compare below.

\subsection{Origin of the \hi\ morphology?}\label{sec:6.2}

HoII is embedded in a massive \hi\ cloud that is significantly more
extended than the optical counterpart. The remarkable morphology of
the cloud in the low density outskirts, compressed on one side with a
cometary appearance on the opposite side (see
Figure~\ref{fig1}), is very suggestive of ram pressure stripping
due to the presence of a hot IGM.  While there are no X-ray
observations to date confirming the presence of diffuse hot gas in the
vicinity of HoII, such observations are available for other much
smaller, `poor' groups \citep[e.g.][]{zab98} in which the hot IGM
density can be as high as $\sim 5 \times 10^{-4}$ atoms cm$^{-3}$
\citep{mul93,sun12}, i.e. about two orders of magnitude higher than
the minimum IGM density necessary to strip the interstellar medium
(ISM) of a low mass dwarf galaxy like HoII
\citepalias[e.g.][]{bur02,mcc07}.

However, the disturbed appearance of the gas cloud may also be the
result of gravitational interactions. While HoII is relatively
isolated from the massive galaxies of the M81 group (see
Figure~\ref{fig9}), it has at least two companion dwarf galaxies
within a projected distance of $\sim$100~kpc which may have tidally
affected the neutral hydrogen cloud.  \citetalias{bur02} discussed the
possible origins of the \hi\ morphology and concluded that the ram
pressure and tidal interaction scenarios were both plausible given the
available information. Their follow-up study, despite the significant
improvement in both coverage and sensitivity, was not sufficient to
rule-out one or the other scenario \citep{bur04}.  They suggested that
deep stellar photometry could resolve the issue by revealing a stellar
counterpart to the northwest \hi\ structure, since ram pressure is not
expected to have an effect on stars.  Here we summarize the current
arguments in favor of each scenario and discuss whether our deep
photometry may help elucidate which is the main mechanism at play.

The deeper \hi\ observations of \citet{bur04} revealed that the low
surface brightness component in the northwest of HoII can be resolved
in two or three arms, which are often associated with tidal
interaction events \citep[e.g.][]{dob10}. As shown in
Section~\ref{sec:4}, the MS+BSG star distribution appears slightly
distorted too, with spiral arm-like features south and west of HoII.
The stellar arms also follow the same counter-clockwise orientation as
the \hi\ arms. However, \citet{bur04} also noted that the arm
kinematics followed the regular rotation of the inner gas, contrary to
what would be expected if they were torn from the main body by tidal
interactions. In addition, while the young stars do seem to trace
spiral arms, the older populations have a much more regular circular
distribution.  Tidal forces would affect all populations equally and
the resulting stellar arms and/or tails would be made of stars of all
ages. This suggests that most of the recent SF occured in the denser
\hi\ arms, resulting in a similar arm-like distribution of the young
stars, rather than stellar tails due to tidal forces. The \hi\ arms,
in turn, could be the consequence of ram pressure, as models which
include gas cooling have shown that it can produce spiral arms
\citep[e.g.][]{sch01,vol03,map08}.

Further insight comes from the detailed SFHs of HoII by \citet{wei08}
and \citet{dal12}, which reveal a significant enhancement in the last
few hundred million years. Could this enhancement also be a
consequence of the process that led to the striking morphology of the
\hi\ cloud? Strong bursts of star formation are often caused by an
external trigger such as an accretion event or interaction with a
nearby galaxy \citep[e.g.][]{ken87,ber12,cig12}. Indeed, as shown in
Figure~\ref{fig9}, HoII has two close companion dwarf galaxies:
M81dwA (=Kar52) located at a projected distance of $\sim$30~kpc to the
North East, and UGC4483, $\sim$110~kpc to the South East). The former,
in particular, is sufficiently close that the timescale of a past
interaction is roughly compatible with the enhanced SFR found by
\citet{dal12}. Assuming a relative velocity of 100~km~s$^{-1}$
\citep[the difference of radial velocities is
$\sim$45~km~s$^{-1}$;][]{kar02}, means that M81dwA could have passed
close to HoII about 300~Myr ago, in good agreement with the beginning
of the SFR enhancement. Both companion dwarfs also appear to have
been experiencing higher star formation in the past few hundred million
years than at older epochs \citep{mcq10,war11}.

However, tidal interaction is not the only process that can lead to
enhanced star formation. In fact, hydrodynamical and N-body
simulations of the interaction between the hot IGM and the ISM of a
galaxy have shown that ram pressure can also result in increased star
formation in the inner regions. The shockwaves resulting from the gas
collision can increase significantly the central gas surface density
and lead to the collapse of molecular clouds, thus enhancing the star
formation rate \citep{qui00,sch01,vol01,mar03}. The fact that the
spatial distribution of young stars closely follows that of the \hi\
contours strongly supports this interpretation. Such an enhancement of
the star formation as a consequence of the interaction between ISM and
IGM has also been observed in various galaxies as asymmetric H$\alpha$
enhancements, usually located along the leading edge of the galaxy
disc \citep[e.g.][]{koo04,cro05}.

Therefore, while the morphology of the \hi\ cloud and the observed
enhancement of the recent star formation rate cannot help discriminate
between the phenomena at play, the regular circular distribution of
the intermediate-age and old stars, the regular rotation of the \hi\
arms, the undisturbed \hi\ distribution of the two nearest dwarf
companions and the lack of \hi\ bridges/filaments between them and
HoII \citep{bur04,chy09,ott12}, all support the ram
pressure stripping scenario. In addition, we do not find a stellar
counterpart to the \hi\ cloud beyond R$\sim$7$\arcmin$ where it
becomes distorted and forms arms/tails, whereas tidal forces would
affect gas and stars equally. Therefore, our data strongly suggest
that ram pressure stripping is the main process responsible for the
swept-back appearance of the \hi\ cloud.

\section{Summary and Conclusions}\label{sec:7}

We have carried out a wide-field survey of the M81 group dwarf galaxy
Holmberg~II based on deep Subaru/Suprime-Cam imaging in $V$ and $I$.
These observations cover the whole extent of the galaxy, including the
vast \hi\ cloud, and allow us to perform photometry of individual
stars down to $I\sim25.2$, i.e. about 1.5~mag below the tip of the
RGB.

The deep CMDs reveal the presence of stellar populations of all ages,
from a few Myr old (MS+BSG, RSG) to several Gyr old (RGB).  While in
most dwarf galaxies in the Local Universe the younger stars are found
to be more centrally concentrated than the older populations
\citep[e.g.][]{zha12}, we find that in HoII the old RGB stars are
significantly more concentrated than the young MS+BSG stars.
Indeed, we find that the exponential scalelength for the young MS+BSG
population is much larger than that of the RGB component
(2.8$\arcmin\pm0.4\arcmin$ vs.\ 0.76$\arcmin\pm0.04\arcmin$,
respectively). We speculate that the shockwave due to ram pressure
increased the gas density in the central part of the \hi\ cloud and
triggered star formation on large scales.

Our Subaru data enable us to construct a composite surface brightness
profile for HoII by combining diffuse light in the central region with
star counts at large radii.  This profile is one of the deepest yet
published for any galaxy, extending from the centre out to ${\rm R}
\sim$~7$\arcmin$ where $\mu_V=32$ mag arcsec$^{-2}$.  Fitting an
exponential profile to the outer regions gives a (projected)
scalelength of $0.70\arcmin\pm0.01\arcmin$, corresponding to $0.69\pm
0.01$~kpc at the distance of HoII.

Finally, we discuss the properties of the resolved stellar populations
in the context of the morphology of the large \hi\ cloud in order to
understand the origin of its swept-back, cometary appearance. Previous
studies based on diffuse optical light or 21~cm data could not
definitively determine whether the cloud shape was due to ram pressure
from a hot IGM or to a tidal interaction with a nearby companion
galaxy.  Our deep photometry shows that the intermediate-age and old
stars have a regular circular distribution and show no sign of tidal
tails/streams. In addition, we find that there are very few, if any,
HoII stars beyond R$\sim$~7$\arcmin$ where the \hi\ becomes distorted.
Since tidal forces would affect gas and stars equally, our data
strongly suggest that the spectacular morphology of the \hi\ cloud is
due to ram pressure. The detection of significant amount of diffuse
hot gas in the vicinity of HoII would further verify this.

\section*{Acknowledgments}

We are very grateful to the anonymous referee for a prompt report that helped us
improve the manuscript, and would like to thank M.\ Bureau for providing the
\hi\ density map of HoII. Support for this work was provided by a rolling grant
from the Science and Technology Facilities Council.  We acknowledge the usage of
the HyperLeda database (http://leda.univ-lyon1.fr). This research has made use
of THINGS, `The \hi\ Nearby Galaxy Survey' \citep{wal08}, and the NASA/IPAC
Infrared Science Archive, which is operated by the Jet Propulsion Laboratory,
California Institute of Technology, under contract with the National Aeronautics
and Space Administration.

\end{document}